\documentclass[twocolumn,10pt]{revtex4-1}
\usepackage{graphicx}
\usepackage[margin=.5in]{geometry}
\usepackage{amsmath}
\usepackage{amssymb}
\usepackage{verbatim}
\usepackage{braket}
\usepackage{float}
\usepackage{breqn}
\usepackage{color}
\usepackage{caption}
\usepackage{subcaption}
\usepackage{multirow}
\makeatletter
\let\cat@comma@active\@empty
\makeatother
\begin{document}
\title{ {\bf    
Physics of emergence beyond Berezinskii-Kosterlitz-Thouless transition for    
interacting topological 
quantum matter 
}
}
\author{Ranjith R Kumar}
\affiliation{ 
Theoretical Sciences Division, Poornaprajna Institute of Scientific Research, Bidalur, 
Bengaluru-562164, India.}
\affiliation{Graduate Studies, Manipal Academy of Higher Education, 
Madhava Nagar, Manipal-576104, India.}
\author{Sujit Sarkar}
\affiliation{Theoretical Sciences Division, Poornaprajna Institute of Scientific Research, 
Bidalur, Bengaluru-562164, India.}
%
\begin{abstract}
\noindent
An attempt is made to find different emergent quantum phases for 
interacting topological state of quantum matter.  
Our study is based on the quantum field theoretical 
renormalization group (RG) calculations. The behaviour of the RG flow 
lines gives the emergence of different quantum phases for non-interacting
and interacting topological state of quantum matter. 
We show explicitly electron-electron interaction can turn a topologically
trivial phase into a non-trivial one and also topological non-trivial phase to
topological trivial phase. 
We show that physics of emergent is go beyond the quantum 
Berezinskii-Kosterlitz-Thouless
transition.
We also present 
the analysis of fixed point and show the behaviour of fixed point
changes in presence and absence of interaction.  
This work provides a new perspective not only from the topological 
state of interacting  quantum matter and but also for the correlated quantum many
body physics.
\end{abstract}

 
\maketitle

{\bf  \large  Introduction}\\
The field of topological state of quantum matter is one of the most active 
research areas in experimental and also in theoretical quantum matter physics $^{1-7}$.
Topological materials are expected to be robust against
disorder $^{8-13}$. 
This makes topological materials candidates
for new electronic devices.
The physics of topological state of quantum matter is typically described
by the band theory and symmetry of non-interacting fermions $^{1-7,9,10}$. 
But in all solids,
electron-electron interaction are unavoidable,
either
screened or weak or sometimes even stronger dominate the physical properties
of a material $^{9,10,14}$ 
electron-electron interaction leads to
the different physical phenomena in quantum many-body systems, such as the Kondo
effect, Mott-Hubbard transition and superconductivity to mention a few $^{9,10,14}$. 
Therefore,
to get a complete picture of the topological state of a quantum many-body system,
one has to consider the
effect of interaction $^{15-31}$.\\
Models of strongly correlated electrons have provided
outstanding challenges to condensed matter theory for many
decades $^{10,14}$.
Most of the previous work on topological superconductivity
in one-dimensional wires has focused on for  
the noninteracting limit $^{1-7,11}$.
Interaction induced topological phases such as topological Kondo insulator,
topological Mott insulator and fractional Chern insulator only exists due
to the interplay of topology and strong correlations $^{11}$.\\
In addition, topological features in strongly correlated
systems are much more readily controllable by external
parameters than in the non-interacting case.
We expect the presence of interaction in the
physical system makes it much more complex than the simple 
non-interacting topological state of quantum matter $^{15-31}$.\\
The physics of one dimensional quantum many body is interesting in its
own right $^{32,33}$.
One dimensional quantum many body system has strong quantum fluctuations that do not
allow spontaneously broken continuous symmetries.
As a result of this, the pairing instabilities do not lead to
any
ordered density-wave $^{10}$. 
This many body has a critical phase with power law decay of various
correlation functions, universally known as a Luttinger liquid $^{10,14}$. 
In one dimensional quantum
many body systems whether it is weakly correlated or strongly proper treatment of the
quantum fluctuations leads in both cases to a Luttinger liquid characterized 
by phonon-like
collective density fluctuation modes. \\ 
In the present study, we use the bosonization process to
recast the model Hamiltonian in continuum field theory.
Our systems can be mapped to a dual-field double sine-Gordon model as a bosonized
effective field theory.
The mathematical structure and results of the RG theory are a significant conceptual
advancement in the quantum field theory of both high-energy and condensed matter physics
$^{34-36}$ in the last several decades. \\ 
{\bf Motivation : } \\
In quantum many-body physics, emergent phenomena are an essential aspect $^{10}$. In this view,
fundamentally new kinds of phenomena emerge in the different region of parameter space
of quantum many body system.
In this paper, we study various quantum emergence phases of one-dimensional 
one-component fermions having proximity-induced pairing
gap and inter particle short-range interaction, as a generalization
of the Kitaev model. 
In this study we raise the question and also solve 
how electron-electron interaction turns a non-trivial
topological state to a topological trivial state and also a topological
non-trivial state to topological trivial state 
for interacting topological quantum matter. 
The detail analysis of fixed point  
for the non-interacting and interacting topological states show many interesting
feature.
The most important part of this study is to show that the emergent quantum phases
are far richer and go beyond the Berizinskii-Kosterlitz-Thouless (BKT) transition $^{37,42}$.\\ 
\begin{figure*}
	\includegraphics[width=12cm,height=7cm]{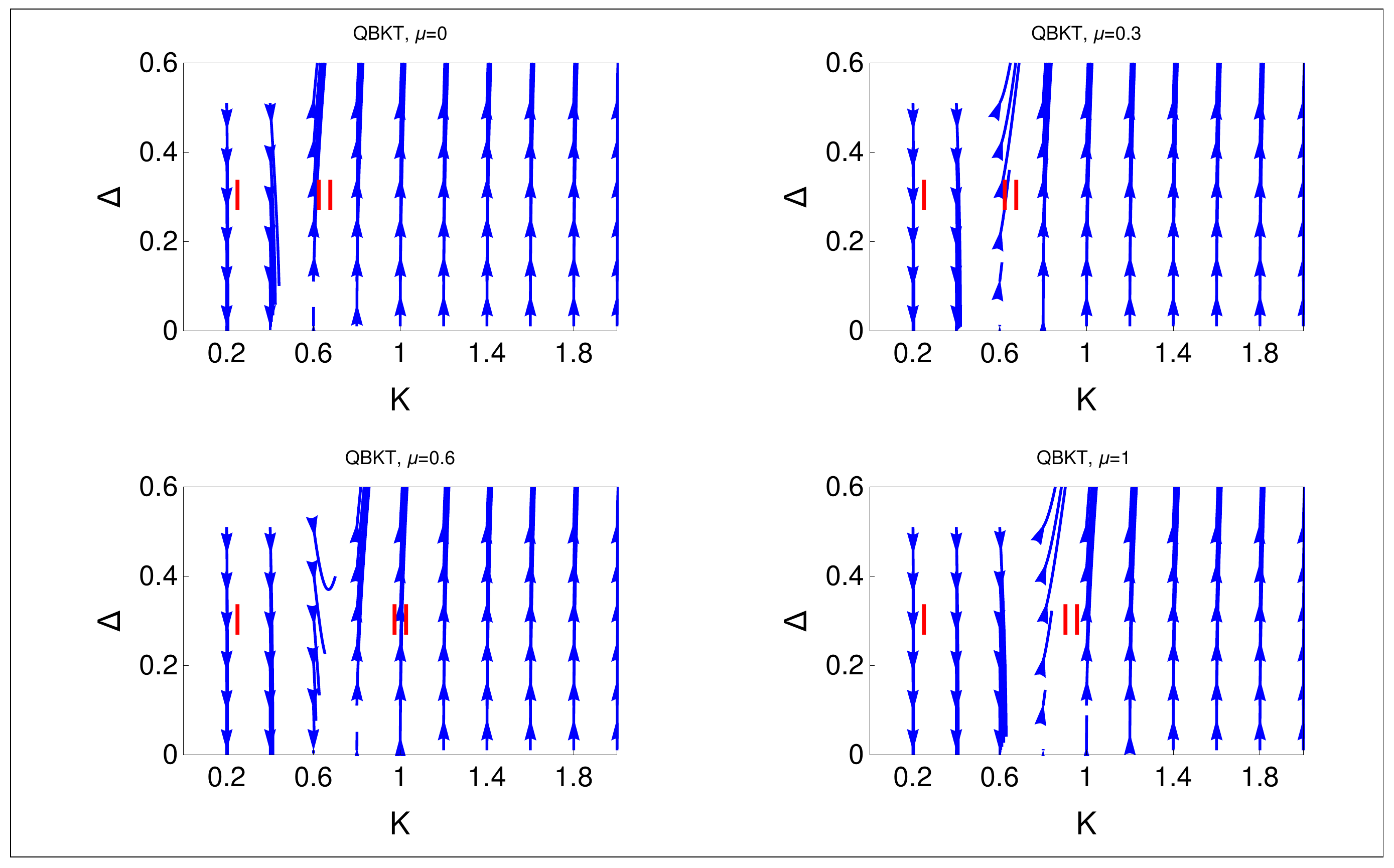}
	\caption{ (Color online.)  
		Behaviour of the RG flow lines for the couplings in the $\Delta$-$K$ plane. 
We 
		present the RG flow lines based on the solution of non-interacting 
RG equations (eq.3).
	}
	\label{Fig. 1 }
\end{figure*}         
\begin{figure*}
	\includegraphics[width=12cm,height=7cm]{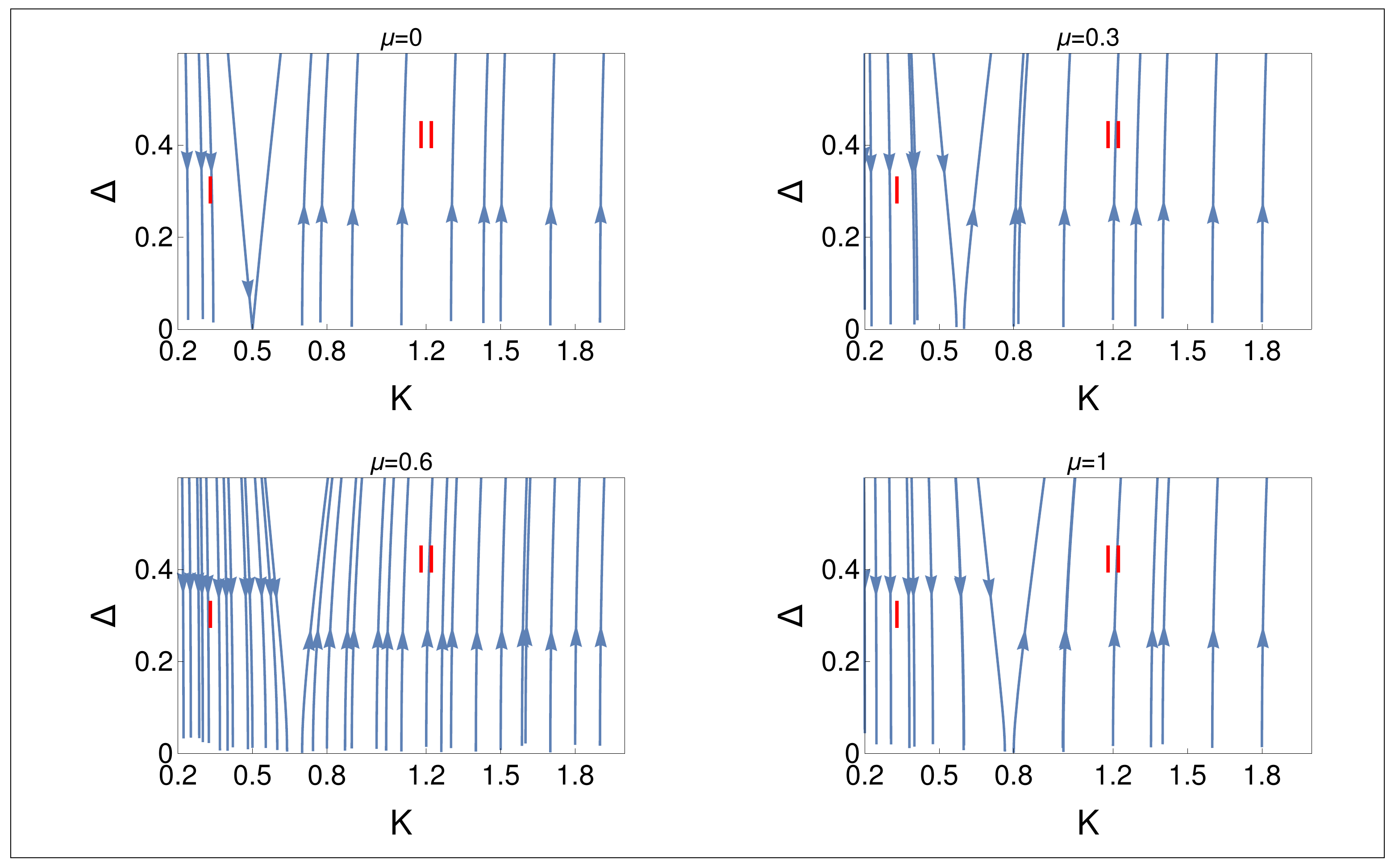}
	\caption{ (Color online.)  
		Behaviour of the RG flow lines for the couplings 
in the $\Delta$-$K$ plane. We 
		present the RG flow lines based on the exact solution (eq. 4).
	}
	\label{Fig. 2 }
\end{figure*}         
\noindent
{\bf  Model Hamiltonians and Renormalization Group Equations for Non-Interacting ($ U=0 $)
Quantum Matter  } \\
The model Hamiltonian for the non-interacting Kitaev chain $^{43,44}$ for the topological state of
quantum matter is the following,
\begin{multline}
H_1 =  - t \sum_{i=1}^{N-1} ( 
{c_i}^{\dagger} {c_{i+1}} + h.c ) 
+ \Delta \sum_{i=1}^{N-1} ( {c_i} {c_{i+1}} + h.c ) \\
-{\mu} \sum_{i}^{N} {c_i}^{\dagger} {c_i} .  
\end{multline}
$t$ is the hopping integral for nearest-neighbour sites, $\Delta$ is the proximity
induced superconducting gap and $\mu $ is the chemical potential. \\
Bosonized version of the model Hamiltonian is the following,
\begin{multline}
H = H_0 + \frac{\Delta}{2}  \int \cos \left( 2 \sqrt{\frac{\pi}{K}}\theta(x)\right)dx +
\mu \sqrt{\frac{K}{\pi}} \int (\partial_x \phi(x)) dx,
\end{multline}
where 
$ H_0 = \frac{v}{2} \int [(\partial_x \theta)^2+ 
(\partial_x \phi)^2] dx $ 
is the non-interacting part of the
Hamiltonian and $v$ is the collective
velocity of the system 
and $K$ is the Luttinger liquid parameter of the system $^{14}$ (detail derivation
is relegated to the "Method" section ). Please notice that initial Hamiltonian
(eq.1) has no $K$ term but the bosonized Hamiltonian (eq.2) has the $K$ term.
We express our model Hamiltonian in terms of two dual fields
$\theta (x)$ and $\phi (x) $, which bosonized the Hamiltonian. 
The fermionic fields for right ($R$) and left ($L$) movers
of a one
dimensional quantum many body system are
$ {\psi}_{R/L, \uparrow/ \downarrow } (x) = \frac{1}{2 \pi \alpha} {\eta_{R, \uparrow}}~
e^{i \sqrt{4 \pi} {\phi}_{R, \uparrow / \downarrow} (x) } $, where
$\eta_{L/R} $ is the Klein factor to preserve the anticommutivity of the
fermionic field which obeys Clifford algebra $^{14}$. 
These two fields are related by the 
relations, $\phi (x) = {\phi}_R (x) + {\phi}_L (x)$ and
$\theta (x) = {\theta}_R (x) + {\theta}_L (x) $. \\
The analytical expressions for the non-interacting RG are the 
following
(detail derivation is relegated to the "Method" section),\\
\begin{align}
\frac{d \Delta}{dl} &= 
\left[  2-\frac{\alpha}{K}
\right ] \Delta  
\nonumber\\ 
\frac{dK}{dl} &=  \frac{{\Delta}^2}{8}  . 
\end{align}
$ \alpha = \left( 1+ \frac{\mu}{\pi \sqrt{\pi}} \right) $, 
( $\alpha =1 $, for $\mu= 0 $).\\
Exact solution for $\Delta $ 
(detail derivation is relegated to the "Method" section),\\
\begin{align}
\Delta = \sqrt{ {{\Delta}_0}^2 + 32 (K - K_0 ) -16 ln(K/K_{0} )
( 1 + \frac{\mu}{\pi \sqrt{\pi}} )}.
\end{align}
{\bf  Model Hamiltonians and Renormalization Group Equations for 
Interacting ($ U \neq 0 $) Quantum Matter } \\
The model Hamiltonian for the interacting Kitaev model is,
\begin{multline}
H_2 =  - t \sum_{i=1}^{N-1} ( 
{c_i}^{\dagger} {c_{i+1}} + h.c ) 
+ \Delta \sum_{i=1}^{N-1} ( {c_i} {c_{i+1}} + h.c ) \\
  + U \sum_{i=1}^{N-1} ( 2 {c_i}^{\dagger} {c_i} - 1)
( 2 {c_{i+1}}^{\dagger} {c_{i+1}} - 1) -{\mu} \sum_{i}^{N} {c_i}^{\dagger} {c_i} .  
\end{multline}
The bosonized form of the interacting Kitaev model Hamiltonian is, 
\begin{multline}
H_2 = H_0 + \frac{\Delta}{2}  
\int \cos \left( 2 \sqrt{\frac{\pi}{K}}\theta(x)\right)dx \\+
U \int \cos \left( 4 \sqrt{\pi K}\phi(x)\right)dx -
\mu \sqrt{\frac{K}{\pi}} \int (\partial_x \phi(x)) dx,
\end{multline}
the third term ($U$) represents the 
intersite repulsive interaction.
In this model Hamiltonian,  
there is no on-site repulsion owing to the Pauli
exclusion principle.\\
The analytical expressions for the interacting RG are the 
(detail derivation is relegated to the "Method" section),\\
\begin{align}
\frac{d \Delta}{dl} &= 
\left[  2-\frac{\alpha}{K}
\right ] \Delta  
+ 2  K \Delta U^2 \nonumber\\ 
\frac{d U}{dl} &=  \left( 2-4K \right) U  \nonumber\\ 
\frac{dK}{dl} &=  \frac{{\Delta}^2}{8} - 8 K^2 U^2 . 
\end{align}
$ \alpha = \left( 1+ \frac{\mu}{\pi \sqrt{\pi}} \right) $, 
( $\alpha =1 $, for $\mu= 0 $).\\
{\bf Results: }\\
{\bf (A). Emergence of Quantum Phases for Non-Interacting Topological State of 
Quantum Matter:  }\\
In fig.1, we present the behaviour of RG flow lines in $\Delta$ -$K$
plane from the solution of RG equations (eq. 3). We study the RG flow lines 
behaviour of coupling with $K$ for the following reasons:
The physics of low-dimensional quantum many body condensed matter system
is enriched
with its new and interesting emergent behavior with the parameter $K$ $^{14}$.
In one dimensional quantum many body system, $K$, plays an important 
role to determine the different emergent quantum phases. 
$K < 1$ and $K > 1$ and $ K=1 $ characterizes
the repulsive, attractive interactions and non-interacting $^{14}$. 
We present four figures for different values of $\mu$.
It reveals from the behaviour of RG flow lines that the upper panel,
the left and right figures are respectively for $\mu =0$ and $\mu =0.3$.
It has only
two quantum phases one is the weak coupling phase, (I), and the other is the
strong coupling phase (II). 
There is no evidence of phase crossover from weak coupling
to strong coupling or vice versa. In the lower panel,   
the left and right figures are respectively for $\mu =0.6$ and $\mu =1$. The
qualitative behaviour of the RG flow lines are the same but the 
gapless LL phase region increase
as we increase $\mu $.\\ 
In fig.2, we present the exact solution for the RG flow equation for the coupling 
$\Delta$ based on
eq.4. This figure panels consists four figures for different values of 
$\mu $ as depicted
in the figures. It is very clear from this figure that for this RG flow diagram system is
in always in two phase regime, i.e., either in the weak coupling phase or in the strong
coupling phase. It reveals from this study that for $\mu =0$ this transition occurs for
$ K=0.5 $. As we increase the value of $\mu $ the transition points shift from $K =0.5$.
This exact solution study is consistent with the numerical study of the RG flow equations
in fig. 1. \\        
{\bf (B). Emergence of Quantum Phases for Interacting Topological State of Quantum Matter:  }\\
Now we present
the result of interacting Kitaev model to show the emergence of different quantum phases in
presence of electron-electron interaction.\\   
\begin{figure}
	\includegraphics[width=\columnwidth,height=6cm]{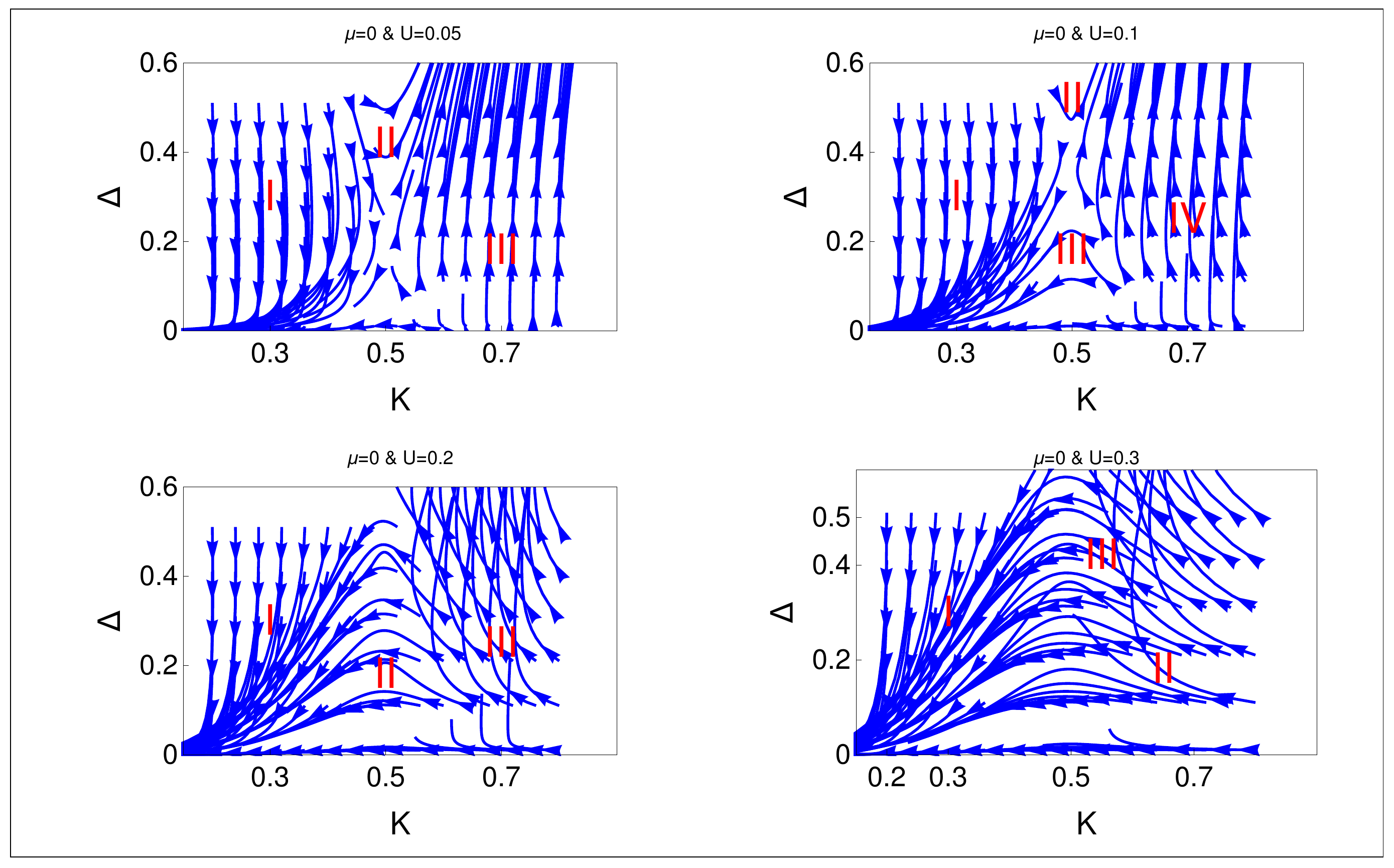}
\caption{  
(Colour online.) Behaviour of the RG flow lines in the $\Delta$- $K$ plane for the
different initial values of $U(0) = (0.05, 0.1, 0.2, 0.3)$. Here we consider $\mu =0$.
We present the RG flow lines from the study of RG equation (eq. 7).  
} 
\label{Fig. 3 }
\end{figure}
\begin{figure}
	\includegraphics[width=\columnwidth,height=6cm]{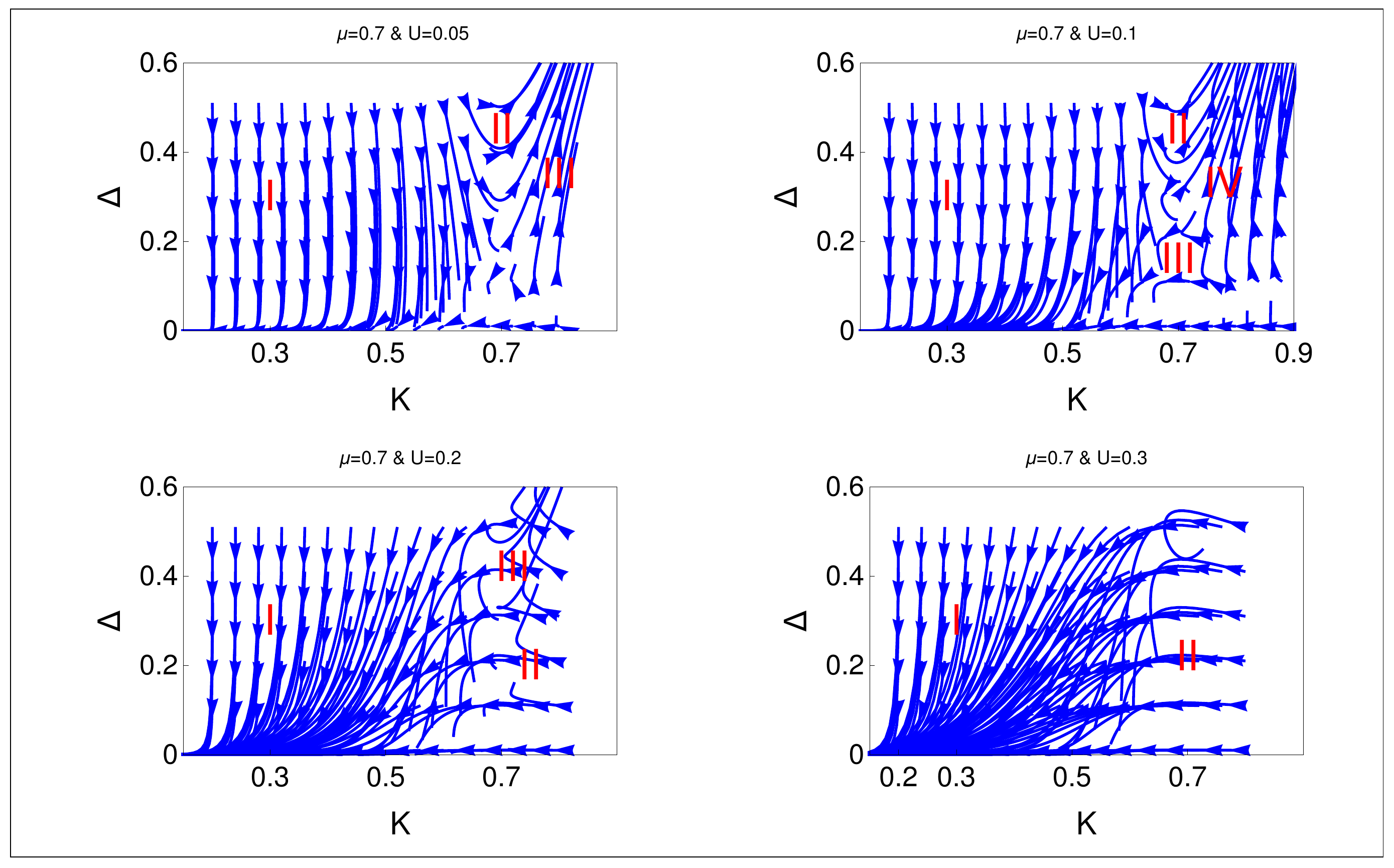}
	\caption{ 
		(Colour online.) Behaviour of the RG flow lines in the $\Delta$- $K$ 
plane for the
		different initial values of $U(0) = (0.05, 0.1, 0.2, 0.3)$. 
Here we consider $\mu =1$.
		We present the RG flow lines from the study of RG equation (eq. 7).  
	} 
	\label{Fig. 4 }
\end{figure}
Fig.3 shows the behaviour of RG flow lines (eq.7) for the coupling $\Delta$ for the 
different initial values of $K$. This figure panel consists of four figures for
the different initial values of $U$ as depicted in the figures and for 
$ \mu =0$. It reveals from the behaviour of the RG flow lines of the
left upper panel, it consists of three different phase regions. The system is
in the weak coupling phase in region (I), where the RG flow lines are flowing off
to the weak coupling phase and finally touch the base line. Region (II) is the phase
crossover region from weak coupling phase to the strong coupling phase, i.e., finally
systems drives to the gapped phase, we term this phase 
crossover as first phase crossover. In region (III), RG flow lines
are flowing off to the strong coupling phase.
\begin{figure}
	\includegraphics[scale=0.3,angle=0]{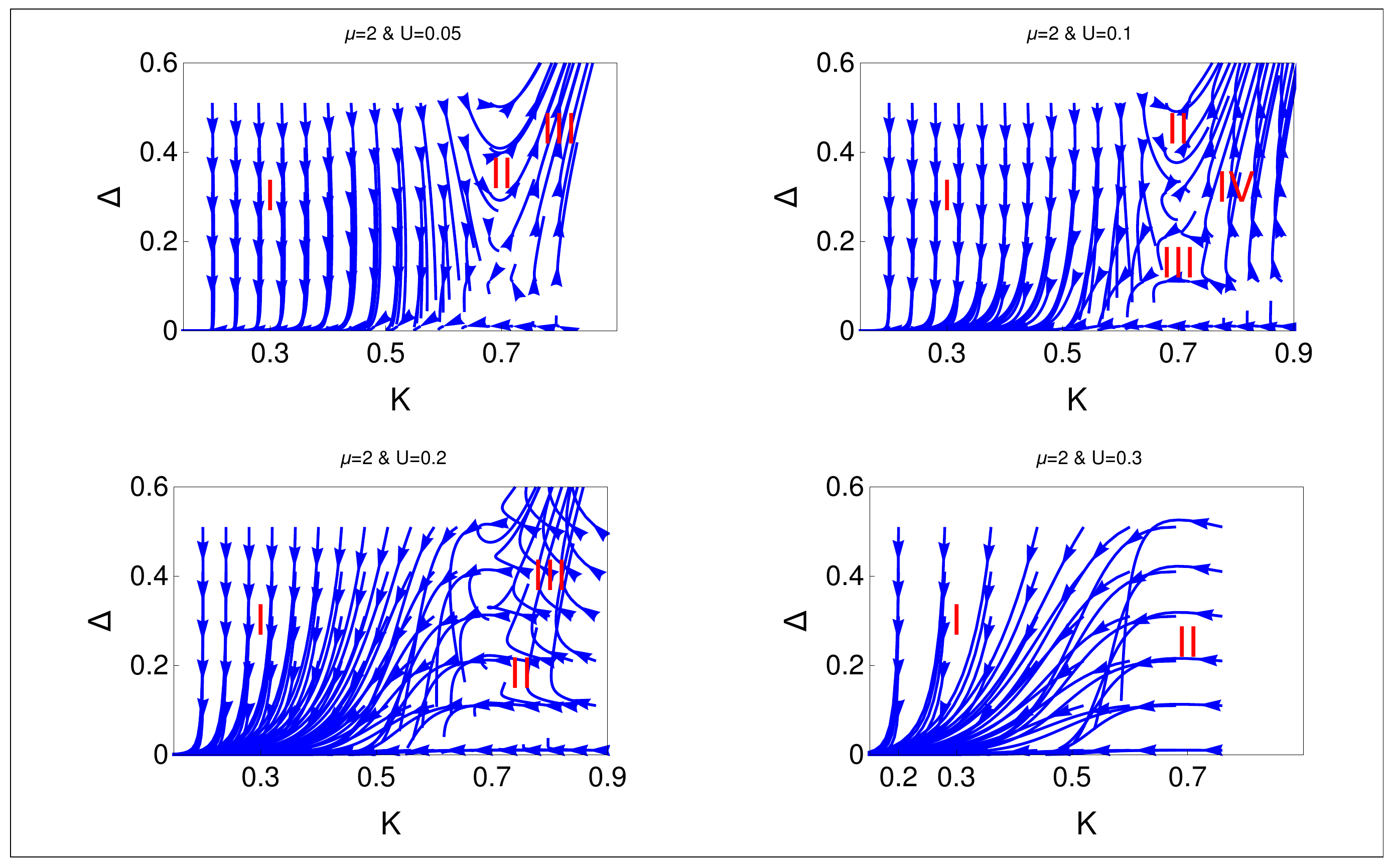}
	\caption{   
		(Colour online.) Behaviour of the RG flow lines in the 
$\Delta$- $K$ plane for the
		different initial values of $U(0) = (0.05, 0.1, 0.2, 0.3)$. 
Here we consider $\mu =0$.
		We present the RG flow lines from the study of RG equation (eq.7).  
	}
	\label{Fig. 5 }
\end{figure}
The right figure of the upper panel is for the higher initial values of 
$U (=0.1) $, the most interesting feature of this RG flow diagram is the appearance
of extra phase crossover from strong coupling to the weak coupling 
due to the reverse flow of the
RG flow lines. We term this phase crossover as second phase crossover (III)
and region (IV) is the strong coupling phase. \\ 
It reveals
from the further higher initial values of $U (=0.2 )$ that the second crossover
phase take the dominant region and the first crossover region totally disappear.
We observe the appearance strong coupling phase for higher initial values of  $K$ 
and $\Delta $.\\ 
The most interesting feature that we observe for further initial
values of $U =0.3$. 
We observe weak coupling phase in region I.
For the higher initial values of $\Delta (\leq 0.22)$, 
the initial phase of
the RG flow lines are strong coupling phase but the direction of these RG flow lines 
change to the weak coupling phase around $ K =0.5$ and sharply  
touches the base line.\\
Another interesting phase we observe from this study is the appearance of a different type
of phase crossover.
We mark this region of phase crossover 
as region (II). In this phase crossover region system drives from flat
phase to the weak coupling phase, we term this phase crossover as a third phase crossover. 
This flat phase occurs for the higher initial values of
$K$ ($ \sim 0.8 $) and for the smaller initial values of $\Delta (\sim 0.2)$. 
The RG flow lines are flowing off with constant values and finally reach the weak 
coupling phase and
touch the base line. 
\begin{figure*}
	\includegraphics[width=12cm,height=7cm]{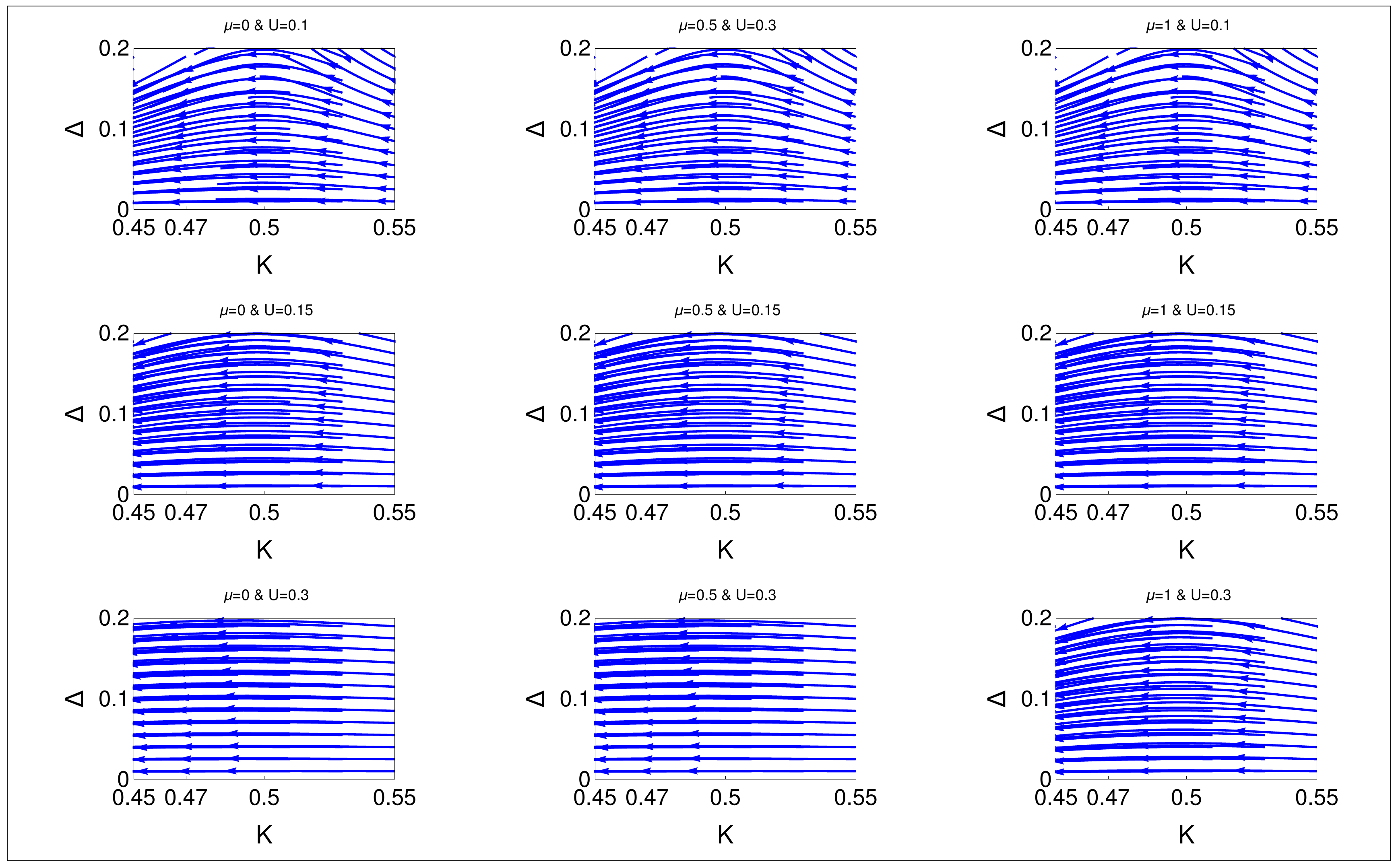}
	\caption{ (Color online.)  
		Behaviour of the RG flow lines for the couplings in the 
$\Delta$-$K$ plane. We 
		present the RG flow lines based on the solution of eq. 7.
	}
	\label{Fig. 6 }
\end{figure*}         
In fig.4 and fig.5, we present the results for $\mu = 0.7$ and $2$ respectively for the 
same RG equation for interacting Kitaev chain. 
The behaviour of  
RG flow lines for the higher values of $\mu$ are differ from the 
$\mu =0$ in the following manner.\\
(1). For $ U= 0.05 $, the phase crossover region shifted to the higher values of $K$. \\ 
(2). For $U =0.1$, the second phase crossover region is very faint for $\mu =0.7$ 
and almost disappear 
for $\mu =2$.\\
(3). Flat phase appears for the smaller initial value of $U =0.2$. \\
(4). For higher initial values of $U=0.3$, we predict only two 
phases and there is no evidence 
of first phase crossover region, as we observe for $\mu=0$. We observe
weak coupling phase and phase crossover from the flat phase region. \\ 
Thus it is clear from
this study how the chemical potential drives to the different emergent phases for
same value of electron-electron interaction.\\ 
Fig.6 consists of three panels. The upper, middle and lower are respectively 
for $U = 0.1, 0.15$ and $0.3$. Each panel consists of three figures for three
different values of $\mu$, the left, middle and right figures are respectively
for $\mu =0 , 0.5$ and
$1$. The main emphasis of this study is to show explicitly the transition from 
the second phase  crossover to the 
flat phase. The upper panel shows the evidence of second phase crossover 
for all values of chemical
potential. We observe that for the higher initial values of $U$, as we notice in the
second and third panel. The second phase crossover has started slowly and finally 
appears as a
flat phase for higher values of $U=0.3$. 
Thus we show explicitly how the electron-electron interaction 
turns the topologically non-trivial state to topologically trivial phase.\\ 
{\bf Physical interpretation of quantum emergence phases and phase crossovers for non-interacting and 
interacting topological quantum states } \\
We present the results for non-interacting and interacting topological state of 
quantum matter through fig.1 to fig. 6. We observe the emergence of three different quantum
phases and three phase crossovers regions. Now we interpret these quantum phase and
phase crossover. Weak coupling phase for the non-interacting system is the gapless
Luttinger liquid phase and the strong coupling phase is the proximity induce topological
superconductng phase (fig.1 and fig.2). \\ For the interacting system, this weak
coupling phase is the charge density wave phase (CDW) phase due to the electron-electron
interaction 
and the strong coupling 
phase is the proximity induce topological superconducting phase. The first phase cross over
region is the transition of the system from CDW phase to the topological superconducting 
phase. The second phase crossover region is the transition from topological superconducting
phase to the CDW phase. Flat phase is for the constant initial value of topological
superconducting phase and it is always associated with the third phase crossover to the
weak coupling CDW phase. There is no evidence of phase crossover from flat phase to
strong coupling phase.\\
It reveals from our study that there are only two phase regions 
for non-interacting phase, there are no phase crossover regions and the
topological superconducting regions shifted with the chemical potential. We have
observed that for interacting phase, three different kind of phase crossover
have appeared as a function of electron-electron interaction and chemical potential. 
Thus the electron-electron
interaction can turn topologically trivial phase to the topologically
non-trivial phase and also the
topologically non-trivial phase to
the topologically trivial phase. 
We have observed that for the higher values of chemical potential
and also for the higher values of $U$, there is no evidence of proximity induced
topological superconducting phase. 
To the best of our knowledge this is the first study in the literature where
the effect of electron-electron interaction 
and the effect of chemical potential has studied 
for topological state of quantum matter rigorously along with the physical interpretation. \\ 
{\bf (C). Characterization of fixed points and stability analysis:}\\
Now we present the nature of fixed points and stability analysis $^{45,46}$ for
the non-interacting and interacting Hamiltonians.\\ 
Stable fixed points: The scaling fields are become irrelevant. These fixed
points are for the stable phase for matter. This fixed point behave as a
attractor.
When one releases the system in the parameter space close to these fixed points
, it scales towards to this fixed point and eventually sits there. These fixed
point is impervious to moderate variations in the microscopic morphology of
the system.\\
Unstable fixed points: The scaling fields are become relevant. These fixed
points are not for the stable phase for matter. The quantum phase of
matter in this fixed point will not stable phase of matter, if the RG
flow lines approach to this fixed point, finally it will be away from it.\\
Marginal fixed points: Marginal scaling field corresponds to a direction in
coupling constant space with vanishing partial derivative. 
For this situation one can consider the second order
derivative ${{\partial}^2 R}|_{g^{\star =0 }}= 2 x $. In the vicinity
of this fixed point, the scaling field then behaves as
$d v_{\alpha} = x {v_{\alpha}}^2 $. The scaling field are marginally
relevant and irrelevant for $x >0$ and $x < 0$ respectively.\\
{\bf Stability matrix for the non-interacting RG equations (eq.3) } \\
Now we do the stability analysis for the fixed point analysis 
of RG equation (3) (detail derivation is relegated to the "Method" section). \\
$\frac{d A_1 }{dl} = B_1 A_1  $,
$A_1 = {( \delta K, \delta \Delta )}^{T} $\\
\begin{equation}
B_1 =  \left(\begin{matrix}
 0 &&
\frac{{\Delta}^{\star}}{4}  \\ 
\frac{\alpha {\Delta}^{\star}}{ {K^{\star}}^2 }  &&
(2 - \frac{\alpha}{ K^{\star} }) \\ 
\end{matrix}\right) ,
\end{equation}
{\bf Stability matrix for the interacting RG equation (eq.7) } \\
$\frac{d A_2 }{dl} = B_2 A_2  $,
$A_2 = {( \delta K , \delta U , \delta \Delta )}^{T} $

\begin{equation}
B_2 =  \left(\begin{matrix}
-16 {{U}^{\star}}^2  K^{\star}&&
-8 {{U}^{\star}}  {K^{\star}}^2 &&
\frac{{\Delta}^{\star}}{4}\\
- 4 {U}^{\star} &&
(2 -4 K) &&
0 \\
C &&
4 {K}^{\star} {\Delta}^{\star} {U}^{\star} &&
D \\
\end{matrix}\right) ,
\end{equation}
where
$ C =  \frac{\alpha}{{ K^{\star}}^2} + 2 {\Delta}^{\star} {{U}^{\star}}^{2} $, 
$ D =  (2 - \frac{\alpha}{ K^{\star}}) + 2 {K}^{\star} {{U}^{\star}}^{2} $, \\
where ${U}^{\star} $, ${\Delta}^{\star} $ and ${K}^{\star } $ are the 
value of $U$, $\Delta$ and $K$ at the fixed point. \\

{\bf Results from the analysis of the fixed point: }\\

{\bf (A). Stability of phase analysis for non-interacting systems} \\
The detail derivation of the analysis of fixed points are relegated in the "Method" section. 
\begin{table*}
	\begin{center}
		\begin{tabular}{ |c|c|c|c| }
			\hline
			K &1/2&1
			&1.5\\
			\hline
			\hline
			$\alpha=1$ 
			& One relevant and one marginal & One relevant and one marginal 
			& All are marginal \\
			
			\hline
			$\alpha=1.5$ & One irrelevant and one marginal coupling 
& One irrelevant and one marginal 
			& One relevant and one marginal\\
			\hline
			$\alpha=2$ & One relevant and one marginal  & All are marginal & 
			One relevant and one marginal \\
			\hline
			\hline
		\end{tabular}
	\end{center}
	\caption{Results of the fixed point analysis for the non-interacting RG equation (eq. 3)
		for $\Delta$ }
	\label{uc}
\end{table*}
\begin{table*}
	\begin{center}	
		\begin{tabular}{ |c|c|c|c| }
			\hline
			K &1/2&1
			&1.5\\
			\hline
			\hline
			$\alpha=1$ 
			&All are marginal & One irrelevant and two marginal 
			& One relevant, one irrelevant  \\
			& & & and one marginal\\
			\hline
			$\alpha=1.5$  &One irrelevant and two marginal  & One relevant, 
one irrelevant  
			& One relevant, one irrelevant \\
			& & and one marginal & and one marginal\\
			\hline
			$\alpha=2$ & One irrelevant and two marginal  & One irrelevant and 
two marginal & 
			One relevant, one irrelevant \\
			& & & and one marginal\\
			\hline
			\hline
		\end{tabular}		
	\end{center}
	\caption{Results of the fixed point analysis for the interacting RG equation (eq. 7)
	}
	\label{uc}
\end{table*}
The analysis of table-I is for the non-interacting RG equation. This table consists of three
rows and three column. Each row is for the different values of $\alpha $ and each column 
is for 
the different values of $K$.\\
It is clear from the non-interacting case for zero chemical potential ($ \alpha =1$ ), 
there is no stable
phase. But for finite chemical potential ($\alpha = 1.5$ ), the system has stable phase for
$K=1/2 $ and $K=1 $ regime but for  
$ K= 1.5$ regime. The system has again no stable system. We find for further
higher values of chemical potential ($\alpha = 2 $) system has no stable phase. \\
{\bf (B). Stability of phase analysis for interacting systems} \\
The analysis of table-II is for the interacting RG equation. This table consists of three
rows and three column. Each row is for the different values of $\alpha $ and each column is 
for 
the different values of $K$.\\ 
It is clear from the interacting case for zero chemical potential ($ \alpha =1$ ), 
there is stable
phase exists only for $K=1$ and $K =1.5$. 
But for finite chemical potential ($\alpha = 1.5$ ), the system has stable phase for the
all values of $K$. We find for further
higher values of chemical potential ($\alpha = 2 $) system has also in the stable phase
for all regime.\\     
{\bf Physics of emergence beyond the quantum Berezinskii-Kosterlitz- Thouless
transition} \\
The BKT mechanism $^{37-42}$, in which a phase transition is mediated 
by the
proliferation of topological defects, governs the critical behavior of a wide range of 
equilibrium 
two dimensional systems with a continuous symmetry, ranging from spin systems to 
superconducting thin films
and two-dimensional Bose fluids, such as liquid helium and ultracold atoms $^{37-42}$. \\
The physics of emergence of quantum phases in low dimensional quantum many body system is
an essential phenomena. Many of the systems have shown the appearance of quantum emergence 
phases and their behaviour is alike to the QBKT transition $^{14,31,46}$.
We now discuss the results of emergence quantum physics which we have obtained in 
this study.\\ 
QBKT system possesses two phase regions, weak coupling and strong coupling, and 
a single phase crossover
region. In weak coupling, system is gapless, and in the strong coupling phase system is in the
gapped phase and the coupling is in the relevant phase in the sense of RG. The phase crossover
region is associated with the transition of the system from weak coupling to the strong coupling
phase. \\
We observe the two phases region from the study of the
behaviour of RG flow lines for non-interacting topological quantum matter, 
one is the weak coupling phase and the other is the strong 
coupling phase, thus the behaviour of RG flow lines is different from the 
behaviour of RG flow lines of QBKT. \\ 
For the interacting topological state of matter, we observe the two phase regions 
and one phase crossover region for the lower
values of electron-electron interaction which is consistent with the QBKT physics.
As we increase the value of $\mu$ and $U$, we observe
the different quantum emergent phases and three phase crossovers.
We observe two phases region for both non-interacting and
interacting topological state of quantum matter but the 
characters are different for non-interacting and interacting
topological state of quantum matter. 
Thus the emergent physics of topological excitation is far more enrich than
the conventional QBKT. \\ 
{\bf Discussions} \\
We have shown that
there are no phase crossover physics for non-interacting topological 
state of quantum matter, this system has only two quantum emergence phase.  
We have shown explicitly that there are three phase crossovers regions and
also three quantum phases for interacting topological state of quantum matter.
We have shown explicitly that how the electron-electron interaction can turn a
topologically trivial state to topologically non-trivial state at the same time 
we have also shown how the electron-electron interaction can turn a non-trivial topological
phase to topologically trivial phase. We have shown the physics of emergence for
interacting topological state of quantum matter is beyond the BKT
transition.  
This work provides a new perspective for the topological 
state of interacting  quantum matter and also for the correlated quantum many
body physics.
\newpage
\onecolumngrid
{\bf Method } \\
{\bf (A). Scaling law and critical exponent }\\
It is well known in the literature that the diverging coherence
length ($ \xi \rightarrow \infty $) is the signature of second
order quantum phase transition. Here we discuss the basic physics
and mathematical analysis of the fixed points and stability
analysis of the RG equations and from that analysis we interfere
about the stable and unstable phases of the system $^{45,46}$. 
\\ Our main
interest is to study the behaviour of the flow in the immediate
vicinity of the fixed point manifolds. One can do the stability
analysis from the following condition. When the coupling constant
$ \lambda$ is close to the fixed point $ {\lambda}^* $. \\
$ R( \lambda ) \equiv  R (\lambda - {\lambda}^*
 + {\lambda}^* ) \simeq W (\lambda -{\lambda}^* ) $, where
$W_{ab} = {( \frac{\partial R_a}{\partial g_b } )|}_{\lambda =
{\lambda}^* } $.\\
To explain the properties of RG flow we consider the
following steps. At first, we diagonalize the matrix $W$
, suppose our coupling parameter space is $W $ then the Eigenvalues
are $ g_1 , g_2 , .........g_N $. The left eigenvector is
$  {\phi}^{\alpha} $ then the eigen value  equation
$ {{\phi}_{\alpha} }^T W =  {{\phi}_{\alpha}}^{T}
{\lambda}_{\alpha} $. \\
The advantage of proceeding via the unconventional set of
left eigenvectors is that it allows us to conventionally
express the flow of the physical constant under renormalization.
\\
Let $v_{\alpha} $ be the $\alpha $th component of the vector,
$ g - {g}^* $ when represented in the basis
$\{ \phi_{\alpha} \} $, we can write
$ v_{\alpha} = { {\phi}_{\alpha}}^{T} (g - {g}^{*} ) $.
These concept display a particularly simple behaviour under
renormalization.
\begin{equation}
\frac{d v_{\alpha} }{d l } = {\phi_{\alpha}}^{T} \frac{d (g - {g}^* )}{d l }
= {\phi_{\alpha}}^{T} W (g - g^* ) = {\lambda}_{\alpha} v_{\alpha} ,
\end{equation}
where $v_{\alpha} $ is the scaling field which changes by the scaling
factor thus $ v_{\alpha} \simeq e^{l \lambda_{\alpha} } $.
For ${\lambda}_{\alpha} > 0 $, the flow is directed away from the
$v_{\alpha} $ is the scaling field which changes by the scaling
factor thus $ v_{\alpha} \simeq e^{l \lambda_{\alpha} } $.
For ${\lambda}_{\alpha} > 0 $, the flow is directed away from the
critical point and this scaling field is the relevant one. When
$\lambda_{\alpha} < 0 $, the flow is attracted by the fixed point
and the scaling field is the irrelevant one. The scaling field
which are invariant under the flow, $ {\lambda}_{\alpha} =0 $ are \\

{\bf (B). Analysis of fixed point for the non-interacting RG equations (eq. 3) } \\


(1). Here we consider $\alpha = 1$ \\

($ K^{\star}$, ${\Delta}^{\star}$) 
= $ (1/2, 0) $ : \\
Eigenvalues and eigenfunctions of the stability matrix analysis are the following:\\
Eigenvalues are $1,0 $, corresponding eigenfunctions are
$(0, 1)$, $(1, 0)$. \\
 
($ K^{\star}$, ${\Delta}^{\star}$) 
= $ (1, 0) $ : \\
Eigenvalues and eigenfunctions of the stability matrix analysis are the following:\\
Eigenvalues are $1.333, 0 $, corresponding eigenfunctions are
$(0, 1)$, $(1, 0)$. \\

($ K^{\star}$, ${\Delta}^{\star}$) 
= $ (1.5, 0) $ : \\
Eigenvalues and eigenfunctions of the stability matrix analysis are the following:\\
Eigenvalues are $0,0 $, corresponding eigenfunctions are
$(0, 1)$, $(-1, 0)$. \\

We consider $\alpha > 1$ ( i.e., $\mu \neq 0$,  ${\Delta}^{\star} =0$ ). 
For this case, we consider the three
different fixed points. \\

(2). Here we consider $\alpha = 1.5 $ \\
($ K^{\star}$, ${\Delta}^{\star}$) 
= $ (1/2, 0) $ : \\
Eigenvalues and eigenfunctions of the stability matrix analysis are the following:\\
Eigenvalues are $-1, 0 $, corresponding eigenfunctions are
$(0, 1)$, $(-1, 0)$. \\

($ K^{\star}$, ${\Delta}^{\star}$) 
= $ (1, 0) $ : \\
Eigenvalues and eigenfunctions of the stability matrix analysis are the following:\\
Eigenvalues are $0.5, 0 $, corresponding eigenfunctions are
$(0, 1)$, $(-1, 0)$. \\

($ K^{\star}$, ${\Delta}^{\star}$) 
= $ (1.5, 0) $ : \\
Eigenvalues and eigenfunctions of the stability matrix analysis are the following:\\
Eigenvalues are $1, 0 $, corresponding eigenfunctions are
$(0, 1)$, $(-1, 0)$. \\

(3). Here we consider $\alpha = 2 $ \\

($ K^{\star}$, ${\Delta}^{\star}$) 
= $ (1/2, 0) $ : \\
Eigenvalues and eigenfunctions of the stability matrix analysis are the following:\\
Eigenvalues are $-2, 0 $, corresponding eigenfunctions are
$(0, 1)$, $(1, 0)$. \\

($ K^{\star}$, ${\Delta}^{\star}$) 
= $ (1, 0) $ : \\
Eigenvalues and eigenfunctions of the stability matrix analysis are the following:\\
Eigenvalues are $0, 0 $, corresponding eigenfunctions are
$(0, 1)$, $(1, 0)$. \\

($ K^{\star}$, ${\Delta}^{\star}$) 
= $ (1.5, 0) $ : \\
Eigenvalues and eigenfunctions of the stability matrix analysis are the following:\\
Eigenvalues are $0.667, 0 $, corresponding eigenfunctions are
$(0, 1)$, $(-1, 0)$. \\

{\bf (B). Stability analysis for the  
fixed point of interacting topological RG equations (eq. 7) } \\

Now we do the fixed point analysis for the whole set of RG equation.
Here we take three different values of $K$, which corresponds to the
three different region of interaction, i.e, strongly correlated regime ($K=0.5$),
non-interacting regime ($K=1$) and attractive regime ($K=1.5 $). We also present the
results three different values of $\mu$ for each $K$. \\ 

(1). At first we consider $\alpha = 1$ ($\mu =0$, $U^{\star} =0$, ${\Delta}^{\star} =0$). \\
For this case, we consider the three
different fixed points. \\
 ($ K^{\star}$, $ U^{\star} $, ${\Delta}^{\star}$) 
= $ (1/2 , 0, 0  ) $ : \\
Eigenvalues and eigenfunctions of the stability matrix analysis are the following:\\
Eigenvalues are $0, 0, 0$, corresponding eigenfunctions are 
$(1, 0, 0)$, $(0, 1, 0)$, $(0, 0, 1)$. Here all the couplings are the marginal, there
is no evidence of stable phase.\\   

 ( $ K^{\star}$, $ U^{\star} $, ${\Delta}^{\star}$) 
= $ (1 , 0, 0  ) $ : \\
Eigenvalues and eigenfunctions of the stability matrix analysis are the following:\\
Eigenvalues are $-2, 1, 0$, corresponding eigenfunctions are
$(0, 1, 0)$, $(0, 0, 1)$, $(1, 0, 0)$.\\
Here we notice that one of the coupling is irrelevant which corresponds to
the stable phase of the system. The other two couplings are the relevant and
marginal. The relevant coupling corresponds to the unstable phase and the marginal
coupling behave as irrelevent coupling when its approach to the fixed point but  
but it behave as relevant coupling when it very close or slightly away from the fixed
point as we have shown through mathematical analysis. Marginal coupling never gives the
stable fixed point. \\ 
 ( $ K^{\star}$, $ U^{\star} $, ${\Delta}^{\star}$) 
= $ (1.5 , 0, 0  ) $ : \\
Eigenvalues and eigenfunctions of the stability matrix analysis are the following:\\
Eigenvalues are $-4, 1.33, 0$, corresponding eigenfunctions are
$(0, 1, 0)$, $(0, 0, 1)$, $(1, 0, 0)$. \\
Here we notice that one of the coupling is irrelevant which corresponds to
the stable phase of the system. The other two couplings are the relevant and
marginal. The relevant coupling corresponds to the unstable phase. There is 
no evidence stable phase for this coupling.\\ 

We consider $\alpha > 1$ ( i.e., $\mu \neq 0$, $ U^{\star} =0 $, ${\Delta}^{\star} =0$ ). 
For this case, we consider the three
different fixed points. \\

(2). Here we consider $\alpha =1.5$, \\

($ K^{\star}$, $ U^{\star} $, ${\Delta}^{\star}$) 
= $ (1/2 , 0, 0  ) $ : \\

Eigenvalues and eigenfunctions of the stability matrix analysis are the following:\\
Eigenvalues are $-1, 0, 0$, corresponding eigenfunctions are 
$(0, 0, 1)$, $(0, 1, 0)$, $(1, 0, 0)$.\\  
Here we notice that one of the coupling is irrelevant which corresponds to
the stable phase of the system. The other two couplings are the 
marginal. The marginal coupling corresponds to the unstable phase. There is 
no evidence critical surface.\\ 

(E). ($ K^{\star}$, $ U^{\star} $, ${\Delta}^{\star}$) 
= $ (1 , 0, 0  ) $ : \\
Eigenvalues and eigenfunctions of the stability matrix analysis are the following:\\
Eigenvalues are $-2, 0.5, 0$, corresponding eigenfunctions are
$(0, 1, 0)$, $(0, 0, 1)$, $(1, 0, 0)$.  
Here we notice that one of the coupling is irrelevant which corresponds to
the stable phase of the system. The another two couplings are the 
relevant and marginal. The relevant coupling corresponds to the unstable phase. There is 
no evidence critical surface.\\ 

(F). ($ K^{\star}$, $ U^{\star} $, ${\Delta}^{\star}$) 
= $ (1.5 , 0, 0  ) $ : \\
Eigenvalues and eigenfunctions of the stability matrix analysis are the following:\\
Eigenvalues are $-4, 1, 0$, corresponding eigenfunctions are
$(0, 1, 0)$, $(0, 0, 1)$, $(1, 0, 0)$. \\
Here we notice that one of the coupling is irrelevant which corresponds to
the stable phase of the system. The other two couplings are the 
relevant and marginal. The marginal coupling corresponds to the unstable phase. There is 
no evidence critical surface.\\ 

Here we consider $\alpha = 2$, \\

(G). ($ K^{\star}$, $ U^{\star} $, ${\Delta}^{\star}$) 
= $ (1/2 , 0, 0  ) $ : \\
Eigenvalues and eigenfunctions of the stability matrix analysis are the following:\\
Eigenvalues are $-2, 0, 0$, corresponding eigenfunctions are 
$(0, 0, 1)$, $(0, 1, 0)$, $(1, 0, 0)$. \\ 
Here we notice that one of the coupling is irrelevant which corresponds to
the stable phase of the system. The other two couplings are the 
marginal couplings. The marginal coupling corresponds to the unstable phase. There is 
no evidence critical surface.\\

(H). ($ K^{\star}$, $ U^{\star} $, ${\Delta}^{\star}$) 
= $ (1 , 0, 0  ) $ : \\
Eigenvalues and eigenfunctions of the stability matrix analysis are the following:\\
Eigenvalues are $-2, 0, 0$, corresponding eigenfunctions are
$(0, 1, 0)$, $(1, 0, 0)$, $(0, 0, 1)$.\\
Here we notice that one of the coupling is irrelevant which corresponds to
the stable phase of the system. The other two couplings are the 
marginal couplings. The marginal coupling corresponds to the unstable phase. There is 
no evidence critical surface.\\ 

(I). ($ K^{\star}$, $ U^{\star} $, ${\Delta}^{\star}$) 
= $ (1.5 , 0, 0  ) $ : \\
Eigenvalues and eigenfunctions of the stability matrix analysis are the following:\\
Eigenvalues are $-4, 0.66, 0$, corresponding eigenfunctions are
$(0, 1, 0)$, $(0, 0, 1)$, $(1, 0, 0)$. \\

{\bf (D). Derivation of bosonized Hamiltonian } 
\begin{align}
H_{1} & =  - t \sum_{i=1}^{N-1} ( {c_i}^{\dagger} {c_{i+1}} + h.c ) 
+ \Delta \sum_{i=1}^{N-1} ( {c_i} {c_{i+1}} + h.c )
( 2 {c_{i+1}}^{\dagger} {c_{i+1}} - 1) -{\mu} \sum_{i}^{N} {c_i}^{\dagger} {c_i} .  
\end{align}

\begin{align}
H_{2} & =  - t \sum_{i=1}^{N-1} ( {c_i}^{\dagger} {c_{i+1}} + h.c ) 
+ \Delta \sum_{i=1}^{N-1} ( {c_i} {c_{i+1}} + h.c ) + U \sum_{i=1}^{N-1} ( 2 {c_i}^{\dagger} {c_i} - 1)
( 2 {c_{i+1}}^{\dagger} {c_{i+1}} - 1) -{\mu} \sum_{i}^{N} {c_i}^{\dagger} {c_i} .  
\end{align}
We recast this model
Hamiltonian in terms of spin-1/2 operators by using the Jordan-Wigner
transformation to connect the spinless fermion operators to the
spin-1/2 operator, which is below.\\
$ {c_j}^{\dagger} = ( {s_j}^{+})  {\Pi}_{l=1}^{j-1} ( {-s_l}^{z} ) $.  
$ {c_j} = ({s_j}^{-}) {\Pi}_{l=1}^{j-1} ( {-s_l}^{z} ) $. \\
After this transformation the Kitaev Hamiltonian become \\  
\begin{equation}
H_1 = \sum_n [(\frac{t+\Delta}{2})S_n^x S_{n+1}^x + \frac{(t-\Delta)}{2} S_n^y S_{n+1}^y 
- \mu  S_n^z].
\end{equation}
\begin{equation}
H_2 = \sum_n [(\frac{t+\Delta}{2})S_n^x S_{n+1}^x + \frac{(t-\Delta)}{2} S_n^y S_{n+1}^y +
U S_n^z S_{n+1}^z - \mu  S_n^z].
\end{equation}
The above two Hamiltonians are free from $K$. 
Therefore, now our main task is to find the analytical 
expression for spin-1/2 operators in terms of in terms of bosonized fields
$\phi$ and $\theta$ and that also show how $K$ appears in the Kitaev model.\\
We present spin operators interms of $\phi$, $\theta$ and $K$ $^{29}$,\\
We present spin operators interms of $\phi$, $\theta$ and $K$ $^{29}$,
\begin{align}
{S_n}^{x} &= [ \cos (2 \sqrt{\pi K} \phi (x) ) + {(-1)}^{n} ] 
\cos (\sqrt{\frac{\pi}{K}} \theta (x) ,\\ 
{S_n}^{y} &= - [  \cos (2 \sqrt{\pi K} \phi (x) ) + {(-1)}^{n} ] 
\sin (\sqrt{\frac{\pi}{K}} \theta (x) , \\
{S_n}^{z} &=  {(-1) }^{n}  \cos (2 \sqrt{\pi K} \phi (x) )  
+ \sqrt{\frac{K}{\pi}} {\partial_x} \phi (x). 
\end{align}
Bosonized version of non-interacting and interacting Hamiltonians  
are the following.
\begin{equation}
H_1 = H_0 + \frac{\Delta}{2}  \int \cos \left( 2 \sqrt{\frac{\pi}{K}}\theta(x)\right)dx 
- \mu \sqrt{\frac{K}{\pi}} \int (\partial_x \phi(x)) dx,
\end{equation}

\begin{equation}
H_2 = H_0 + \frac{\Delta}{2}  \int \cos \left( 2 \sqrt{\frac{\pi}{K}}\theta(x)\right)dx +
U \int \cos \left( 4 \sqrt{\pi K}\phi(x)\right)dx -
\mu \sqrt{\frac{K}{\pi}} \int (\partial_x \phi(x)) dx,
\end{equation}

$ H_0 = \frac{v}{2} \int [(\partial_x \theta)^2+ 
(\partial_x \phi)^2] dx $. We notice that $H_0$ appears 
with out $K$, therefore the rest three terms of eq. 17 appear as a function
of $K$ otherwise it appears with out $K$. \\ 
{\bf (E). Derivation of exact solutions for the non-interacting RG equation (eq. 3) } \\
We may write the equation for $\frac{d \Delta}{dK}$ from eq. 9 as \\
$$ \frac{d \Delta}{d K} = 
\frac{ 8 [2 - \frac{1}{K} (1 + \frac{\mu}{\pi \sqrt{\pi}})]}{\Delta} $$ \\
We can write the integration constant as \\
$$ C = {\frac{1}{2} {\Delta}^2 } -16 K 
+  8 (1 + \frac{\mu}{\sqrt{\pi}}) ln (K) $$. \\
One can evaluate the constant from the 
initial value of $\Delta$ and $K$, i.e.,
${\Delta}_0 $ and $K_0 $. \\
$$ \Delta =\sqrt{ { {\Delta_0}^2 } + 32 (K-K_0)   
- 16 (1 + \frac{\mu}{\pi \sqrt{\pi}} ) ln (K/{K_0}) } $$. \\
Finally we obtain the exact solution for RG flow lines by integrating 
the RG equations. \\

{\bf (F). Derivation of renormalization group equations for 
non-interacting and interacting Hamiltonians :}\\
Our starting point is the bosonized Hamiltonian,
\begin{equation}
H = H_0 + a \int \cos \left( 2 \sqrt{\frac{\pi}{K}}\theta(x)\right)dx +
\mu \sqrt{\frac{K}{\pi}} \int (\partial_x \phi(x)) dx, 
\end{equation}
where $ H_0 = \frac{v}{2} \int [(\partial_x \theta)^2+ (\partial_x \phi)^2] dx$.
Here, we consider ${\Delta}=2a $ for the smoothness of calculation, 
but we finally present the RG equations in terms of $\Delta$. 
Now we write the partition function $\mathcal{Z}$ in terms of fields as, 
\begin{equation}
\mathcal{Z}= \int \mathcal{D}\phi \mathcal{D}\theta e^{-S_E[\theta]},
\end{equation}
where $S_E$ is the Euclidean action which can be written as
$S_E= -\int dr \mathcal{L} = -\int dr (\mathcal{L}_0 + \mathcal{L}_{int})$, where $r=(\tau,x)$.
Now we divide the fields into slow and fast modes and integrate out the fast modes.
$\theta$ is $\theta(r)=\theta_s(r)+\theta_f(r)$, where
\begin{equation}
\theta_s(r)=\int_{-\Lambda/b}^{\Lambda/b}
\frac{d\omega}{2\pi} e^{-i\omega r}\theta(\omega) \;\;\;\;\&\;\;\;
\theta_f(r)=\int_{\Lambda/b<|\omega_n|<\Lambda} \frac{d\omega}{2\pi}
e^{-i\omega r}\theta(\omega).
\end{equation}
Here $\Lambda$ is the cut-off to start with and $b$ is a factor greater than one.
It is clear from the above definitions of faster and slower mode. One can
make the average over the fast mode in order to get an effective action for 
the slower mode.
Thus $\mathcal{Z}$ is
\begin{equation}
\mathcal{Z}=\int  \mathcal{D}\theta_s
\mathcal{D}\theta_f e^{-S_s(\phi_s,\theta_s)} e^{-S_f(\phi_f,\theta_f)}
e^{-S_{int}(\phi,\theta)}.
\end{equation}
Using the relation $\left\langle A \right\rangle_f =
\int \mathcal{D}\theta_f e^{-S_f(\theta_f)} A $, one can write
\begin{equation}
\mathcal{Z}=\int  \mathcal{D}\theta_s e^{-S_s(\theta_s)}
\left\langle e^{-S_{int}(\theta)} \right\rangle_f.
\end{equation}
We write the effective action as
\begin{equation}
e^{-S_{eff}(\theta_s)}=e^{-S_s(\theta_s)}
\left\langle e^{-S_{int}(\theta)} \right\rangle_f.
\end{equation}
Taking $\ln$ on both sides give,
$ S_{eff}(\theta_s)= S_s(\theta_s) -
\ln\left\langle e^{-S_{int}(\theta)} \right\rangle_f. $
By writing the cumulant expansion up to third order, we have
\begin{dmath}
S_{eff}(\theta_s) = S_s(\theta_s)+
\left\langle S_{int}(\theta) \right\rangle_f -
\frac{1}{2} \left( \left\langle S^2_{int}(\theta) \right\rangle_f -
\left\langle S_{int}(\theta) \right\rangle^2_f \right) +  \\
 \frac{1}{6} \left( \left\langle S^3_{int}(\theta) \right\rangle_f -
3 \left\langle  S^2_{int}(\theta) \right\rangle_f 
\left\langle  S_{int}(\theta) \right\rangle_f  
+ 2 \left\langle S_{int}(\theta) \right\rangle^3_f \right). 
\end{dmath}
At first we derive the 2nd order RG equations for $\mu =0$ and
then we extend it for finite $\mu$. \\
\begin{dmath}
\left\langle S_{int}(\theta)\right\rangle =
\int dr \left[ a\left\langle  \cos \left( 2 \sqrt{\pi} \theta(r)
\right)\right\rangle_f \right] 
\end{dmath}
\begin{dmath}
\int dr \left[ a\left\langle  \cos \left( 2 \sqrt{\pi}
\theta(r)\right)\right\rangle_f \right] = b^{-\frac{1}{K}}
\int dr \left[ a \cos \left( 2 \sqrt{\pi} \theta_s(r)\right) \right] \label{19.pdf.11}
\end{dmath}
Now we calculate the second order cumulant term of the action $S_{eff} (\theta) $
(eq. 25).
\begin{dmath}
-\frac{1}{2}(\left\langle S_{int}^2\right\rangle-\left\langle S_{int}\right\rangle^2)
= -\frac{1}{2} \int dr dr^{\prime} \left\lbrace a^2 [....] 
 \right\rbrace ,
\end{dmath}
where the dotted term represents the expectation value of the correlation function of 
sine-Gordon operators, which
we evaluate below.\\
\begin{dmath}
-\frac{a^2}{2} \int dr dr^{\prime} \left\lbrace \left\langle \cos[2\sqrt{\pi}\theta(r)]
\cos[2\sqrt{\pi}\theta(r^{\prime})] \right\rangle -
\left\langle \cos[2\sqrt{\pi}\theta(r)] \right\rangle  \left\langle
\cos[2\sqrt{\pi}\theta(r^{\prime})] \right\rangle \right\rbrace
= \frac{a^2}{4} \left( 1-b^{-\frac{2}{K}} \right)
\int dr (\partial_r \theta_s(r))^2. \label{19.pdf.21}
\end{dmath}
We obtain the following relation by comparison of rescaled $a$ term 
using the rescaled relation as $b = e^{dl}$.   
\begin{dmath*}
\bar{a}= a b^{(2-\frac{1}{K})}
\end{dmath*}
Finally we write the RG equation in terms of $\Delta (=2a)$ as,
\begin{dmath*}
\frac{d \Delta}{dl}= \left( 2-\frac{1}{K}\right) \Delta 
\end{dmath*}
Comparison of rescaled $K$ terms from the contribution of $\phi$, gives,\\
\begin{dmath*}
\bar{K}= K + \frac{Ka^2}{4}\left( b^2 - b^{(2-\frac{2}{K})}\right)
\end{dmath*}
We obtain the final form of $\frac{dK}{dl}$ after the  
contribution from $\theta$ for $K$. 
\begin{dmath*}
\dfrac{dK}{dl}= \frac{\Delta^2}{8}
\end{dmath*}
Similarly one can find the complete RG equation 
in the following 
way.\\ 
Now we calculate the third order terms of cumulant expansion for the effective
action (eq. 25),
\begin{dmath*}
\frac{1}{6}\left( \left\langle S_{int}^3 \right\rangle -3\left\langle
S_{int}^2 \right\rangle\left\langle S_{int} \right\rangle +2 \left\langle S_{int}
\right\rangle^3\right) = \frac{1}{6} \int d\tau d\tau^{\prime} d\tau^{\prime \prime}
\\\left\lbrace
 a^3 \left[ ..... \right] + U^3 \left[ ..... \right]
+ a^2 U \left[ ..... \right] + a U^2 \left[ ..... \right]
\right\rbrace
\end{dmath*}
\begin{dmath}
\frac{a^3}{6} \int d\tau d\tau^{\prime} d\tau^{\prime \prime} 
\left[ \left\langle \cos[\sqrt{4\pi}\theta (\tau)] 
\cos[\sqrt{4\pi}\theta (\tau^{\prime})] \cos[\sqrt{4\pi}\theta (\tau^{\prime \prime})] 
\right\rangle - 3 \left\langle \cos[\sqrt{4\pi}\theta (\tau)] \cos[\sqrt{4\pi}\theta 
(\tau^{\prime})] \right\rangle \left\langle \cos[\sqrt{4\pi}\theta (\tau^{\prime \prime})] 
\right\rangle + 2 \left\langle \cos[\sqrt{4\pi}\theta (\tau)] \right\rangle \left\langle 
\cos[\sqrt{4\pi}\theta (\tau^{\prime})] \right\rangle \left\langle 
\cos[\sqrt{4\pi}\theta (\tau^{\prime \prime})] \right\rangle   \right] = \\
 \left( \frac{a^3}{24}\right) (6b^{-\frac{3}{K}}-3b^{-\frac{5}{K}}-3b^{-\frac{1}{K}}) 
\int d\tau  \cos[\sqrt{4\pi}(\theta_s(\tau))]. \label{mu8}
\end{dmath}
Following the same procedure used to derive the second order RG,
we obtain the following equations.\\
\begin{dmath*}
da=  \left( 2-\frac{1}{K}\right)a dl 
\end{dmath*}
Finally we write this equation in term of $\Delta $,
\begin{dmath*}
\frac{d \Delta}{dl}= \left( 2-\frac{1}{K}\right) \Delta 
\end{dmath*}
We obtain, by comparing the $\theta$ terms, 
\begin{equation}
\frac{dK}{dl}=  \frac{a^2}{2}= \frac{\Delta^2 }{8}
\end{equation}
{\bf Modified RG equations in presence of $\mu$ }\\
Now we derive the RG equations for finite chemical potential ($\mu \neq 0$).
The presence of finite chemical potential yields five extra terms
in the second order cumulant expansion (eq. 29), of which three are vanishing. The
non-vanishing contributions are from the two correlation functions,
(a). $\int dr dr^{'} a i \frac{\mu}{\sqrt{\pi}}$, and
(b). $\int dr dr^{'} i \frac{\mu} {\sqrt{\pi}} a $, \\
Now we calculate $ a i \frac{\mu}{\sqrt{\pi}}$ term:
\begin{dmath*}
-\frac{1}{2} \int d\tau d\tau^{\prime}\left(- \frac{a i}{\mu} \sqrt{\pi}
\left\langle \cos(\sqrt{4\pi}\theta(\tau))
\partial_{\tau^{\prime}}\theta (\tau^{\prime})\right\rangle
- \left\langle \cos(\sqrt{4\pi}\theta(\tau)) \right\rangle \left\langle
\partial_{\tau^{\prime}}\theta (\tau^{\prime})\right\rangle \right)
= \frac{a i \mu}{2 \sqrt{\pi}}\int d\tau d\tau^{\prime}
\left[ \left\langle \cos[\sqrt{4\pi}\theta(\tau)]
(\partial_{\tau^{\prime}}\theta_s(\tau^{\prime})
+\partial_{\tau^{\prime}}\theta_f(\tau^{\prime}))\right\rangle
- \left\langle \cos[\sqrt{4\pi}\theta(\tau)]
\right\rangle \left\langle \partial_{\tau^{\prime}}\theta_s(\tau^{\prime})
+\partial_{\tau^{\prime}}\theta_f(\tau^{\prime}) \right\rangle \right] \\
=\frac{a i \mu}{2 \sqrt{\pi}}\int d\tau d\tau^{\prime}\left[\left\langle
\cos[\sqrt{4\pi}\theta(\tau)]\partial_{\tau^{\prime}}
\theta_f(\tau^{\prime})\right\rangle \right] \\
=-\frac{a \mu}{\sqrt{\pi} 2\pi} (1-b^{-\frac{1}{K}})\int d\tau \cos[\sqrt{4\pi}\theta_s(\tau)]
\end{dmath*}
Similarly one can calculate the,
$\int dr dr^{'} \frac{i \mu}{ \sqrt{\pi}} a $. \\ Finally, combination of these
two terms ($a+b$) yields
$ -\frac{a \mu }{\pi \sqrt{\pi}}
(1-b^{-\frac{1}{K}})\int d\tau \cos[\sqrt{4\pi}\theta_s(\tau)$, as a consequence of it
$a$ term modify to
\begin{dmath*}
\bar{a} =a\left( b^{2-\frac{1}{K}}\right) -\frac{ a \mu }{ \pi \sqrt{\pi}} (b^2-b^{2-\frac{1}{K}}) .
\end{dmath*}
Finally we obtain the modified 2nd order and 3rd RG equations in presence of
$\mu$ from the rescaled $a$ term as \\
\begin{dmath*}
\bar{a} =a \left( b^{2-\frac{1}{K}}\right) -
\frac{ a \mu }{\pi \sqrt{\pi}} (b^2-b^{2-\frac{1}{K}})
+ \left( \frac{a^3}{24}\right) (6b^{2-\frac{3}{K}}-3b^{2-\frac{5}{K}}-3b^{2-\frac{1}{K}})
\end{dmath*}
The correlation functions for finite $\mu$, only show up for the RG equations of
$\frac{da}{dl}$ because only these two correlation functions give non vanishing 
contributions.
The 3rd order RG equation in terms of $\Delta$ as
\begin{dmath}
\frac{d \Delta}{dl}= \left[  2-\frac{1}{K}\left
( 1+ \frac{\mu}{\sqrt{\pi}}\right)\right] \Delta 
\end{dmath}

\vspace{1cm}
{\bf Acknowledgments}
The author would like to acknowledge Dr. R. Srikanth  
for reading the manuscript critically. 
The author also 
would like to acknowledge  
RRI library, DST (EMR/2017/000898) 
for books/journals.\\\\ 
\textbf{Author contributions statement:}\\
S.S. identified the problem and also write the manuscript, R.R.K. do the calculations of this problem
under the guidence of S.S.
All authors have analysed the results and reviewed the manuscript. \\
{\bf Additional Informations: } \\
Competing interests: The author declare no competing interests.  \\
  

\begin{thebibliography}{10}
\bibitem{hasan1} 
Hasan, M. Zahid, and Charles L. Kane. "Colloquium: topological insulators." Reviews of modern physics 82, no. 4 (2010): 3045.
\bibitem{hasan2} 
Hasan, M. Zahid, and Joel E. Moore. "Three-dimensional topological insulators." Annu. Rev. Condens. Matter Phys. 2, no. 1 (2011): 55-78.
\bibitem{zhang1} 
Maciejko, Joseph, Taylor L. Hughes, and Shou-Cheng Zhang. "The quantum spin Hall effect." Annu. Rev. Condens. Matter Phys. 2, no. 1 (2011): 31-53.
\bibitem{zhang2} 
Yan, Binghai, and Shou-Cheng Zhang. "Topological materials." Reports on Progress in Physics 75, no. 9 (2012): 096501.
\bibitem{berni1} 
Bernevig, B. A., and T. L. Hughes. "Topological Insulators and Topological Superconductors" (Princeton, NJ: Princeton University Press)
(2013).
\bibitem{zhang3} 
Zhang, Haijun, and Shou‐Cheng Zhang. "Back Cover: Topological insulators from the perspective of first‐principles calculations". Phys. Status Solidi RRL 1–2 (2013).
\bibitem{ren} 
Ren, Yafei, Zhenhua Qiao, and Qian Niu. "Topological phases in two-dimensional materials: a review." Reports on Progress in Physics 79, no. 6 (2016): 066501.
\bibitem{wil} 
Wilczek, Frank. "Majorana returns." Nature Physics 5, no. 9 (2009): 614-618.
\bibitem{frad} 
Fradkin, E. Field Theories in Condensed Matter Physics (Cambridge
University Press, New Delhi, 2013).
\bibitem{girvin} 
Girvin, S., and Yang, K. Modern Condensed Matter Physics (Cambridge
University Physics, New Delhi, 2019).
\bibitem{rac1}
Rachel, Stephan. "Interacting topological insulators: a review." 
Reports on Progress in Physics 81, no. 11 (2018): 116501.
\bibitem{suj1} 
Sarkar, Sujit. "Physics of Majorana modes in interacting helical liquid." 
Scientific reports 6, no. 1 (2016): 1-7.
\bibitem{beena} 
Beenakker, C. W. J. "Search for Majorana fermions in superconductors." Annu. Rev. Condens. Matter Phys. 4, no. 1 (2013): 113-136.
\bibitem{gia1} 
Giamarchi, T. Quantum Physics in One Dimension 
(Clarendon Press, Oxford, 2003).
\bibitem{fu1} 
Fu, Liang, and Charles L. Kane. "Superconducting proximity effect and Majorana fermions at the surface of a topological insulator." Physical review letters 100, no. 9 (2008): 096407.
\bibitem{qi} 
Qi, Xiao-Liang, and Shou-Cheng Zhang. "The quantum spin Hall effect and topological insulators." arXiv preprint arXiv:1001.1602 (2010).
\bibitem{qi2}
Qi, Xiao-Liang, and Shou-Cheng Zhang. "Topological insulators and superconductors." Reviews of Modern Physics 83, no. 4 (2011): 1057.
\bibitem{kane} 
Kane, Charles L., and Eugene J. Mele. "Quantum spin Hall effect in graphene." Physical review letters 95, no. 22 (2005): 226801.
\bibitem{hal}
Haldane, F. Duncan M. "Model for a quantum Hall effect without Landau levels: Condensed-matter realization of the" parity anomaly"." Physical review letters 61, no. 18 (1988): 2015.
\bibitem{int1} 
Nadj-Perge, Stevan, Ilya K. Drozdov, Jian Li, Hua Chen, Sangjun Jeon, Jungpil Seo, Allan H. MacDonald, B. Andrei Bernevig, and Ali Yazdani. "Observation of Majorana fermions in ferromagnetic atomic chains on a superconductor." Science 346, no. 6209 (2014): 602-607.
\bibitem{int2} 
Dumitrescu, Eugene, Brenden Roberts, Sumanta Tewari, Jay D. Sau, and S. Das Sarma. "Majorana fermions in chiral topological ferromagnetic nanowires." Physical Review B 91, no. 9 (2015): 094505.
\bibitem{int3a} 
Li, Zhidan, and Qiang Han. "Effect of interaction on the Majorana zero modes in the Kitaev chain at half filling." Chinese Physics Letters 35, no. 4 (2018): 047101.
\bibitem{int4} 
Chan, Y-H., Ching-Kai Chiu, and Kuei Sun. "Multiple signatures of topological transitions for interacting fermions in chain lattices." Physical Review B 92, no. 10 (2015): 104514.
\bibitem{int5} 
Katsura, Hosho, Dirk Schuricht, and Masahiro Takahashi. "Exact ground states and topological order in interacting Kitaev/Majorana chains." Physical Review B 92, no. 11 (2015): 115137.
\bibitem{fid1}
Fidkowski, Lukasz, and Alexei Kitaev. "Effects of interactions on the topological classification of free fermion systems." Physical Review B 81, no. 13 (2010): 134509.
\bibitem{fid2}
Fidkowski, Lukasz, and Alexei Kitaev. "Topological phases of fermions in one dimension." Physical review b 83, no. 7 (2011): 075103.
\bibitem{int6} 
Rahmani, Armin, Xiaoyu Zhu, Marcel Franz, and Ian Affleck. "Emergent supersymmetry from strongly interacting Majorana zero modes." Physical review letters 115, no. 16 (2015): 166401.
\bibitem{int7} 
Gergs, Niklas M., Lars Fritz, and Dirk Schuricht. "Topological order in the Kitaev/Majorana chain in the presence of disorder and interactions." Physical Review B 93, no. 7 (2016): 075129.
\bibitem{suhas} 
Lutchyn, Roman M., and Matthew PA Fisher. "Interacting topological phases in multiband nanowires." Physical Review B 84, no. 21 (2011): 214528.;
Gangadharaiah, Suhas, Bernd Braunecker, Pascal Simon, and Daniel Loss. 
"Majorana edge states in interacting one-dimensional systems." Physical 
review letters 107, no. 3 (2011): 036801.
\bibitem{oleg} 
Stoudenmire, E. M., Jason Alicea, Oleg A. Starykh, and Matthew PA Fisher. 
"Interaction effects in topological superconducting wires supporting Majorana 
fermions." Physical Review B 84, no. 1 (2011): 014503.
\bibitem{suj3}
Sarkar, Sujit.,
" A Study of Interaction Effects and
Quantum Berezinskii- Kosterlitz-
Thouless Transition in the Kitaev
Chain "
Scientific reports {\bf 10}, 2299 (2020).
\bibitem{hohen} 
Hohenberg, Pierre C. "Existence of long-range order in one and two dimensions." 
Physical Review 158, no. 2 (1967): 383.
\bibitem{mermin} 
Mermin, N. David, and Herbert Wagner. "Absence of ferromagnetism or antiferromagnetism in one-or 
two-dimensional isotropic Heisenberg models." Physical Review Letters 17, no. 22 (1966): 1133.
\bibitem{zee}
Zee, A. Quantum Field Theory in a NutShell, Universities Press, Hyderabad (2013)
\bibitem{shankar1} 
Shankar, Rev. "Renormalization-group approach to interacting fermions." 
Reviews of Modern Physics 66, no. 1 (1994): 129.
\bibitem{shankar2} 
Shankar, R. "Quantum Field Theory and Condensed Matter: An 
Introduction" (Cambridge University Press, New Delhi, 2017).
\bibitem{ortiz} 
Ortiz, G., E. Cobanera, and Z. Nussinov. "Berezinskii–Kosterlitz–Thouless transition through the eyes of duality." In 40 Years of Berezinskii–Kosterlitz–Thouless Theory, pp. 93-134. 2013.
\bibitem{bere} 
Berezinskii, V. L. "Destruction of long-range order in one-dimensional and two-dimensional systems possessing a continuous symmetry group. II. Quantum systems." Sov. Phys. JETP 34, no. 3 (1972): 610-616.
\bibitem{koster} 
Kosterlitz, John Michael, and David James Thouless. "Ordering, metastability and phase transitions in two-dimensional systems." Journal of Physics C: Solid State Physics 6, no. 7 (1973): 1181.
\bibitem{dalibard}
Hadzibabic, Zoran, Peter Krüger, Marc Cheneau, Baptiste Battelier, and Jean Dalibard. "Berezinskii–Kosterlitz–Thouless crossover in a trapped atomic gas." Nature 441, no. 7097 (2006): 1118-1121.
\bibitem{lec} 
Haldane. D., (Nobel Prize in Physics 2016), Distinguished lecture on 11 January
2019 at ICTS, India.
\bibitem{jose}
José, Jorge V. "Duality, gauge symmetries, renormalization groups and the BKT Transition." International Journal of Modern Physics B 31, no. 6 (2017): 1730001.
\bibitem{majo1} 
Majorana, Ettore. "Teoria simmetrica dell’elettrone e del positrone." 
Il Nuovo Cimento (1924-1942) 14, no. 4 (1937): 171-184.
\bibitem{kitaev} 
Kitaev, A. Yu. "Unpaired Majorana fermions in quantum wires." Physics-uspekhi 44, no. 
10S (2001): 131.
\bibitem{kerson} 
Huang, K., Statistical Mechanics (
John Wiley and Sons, Singapore, 1976).
\bibitem{Altland} 
Altland, A., and Simons, B. Condensed Matter Field Theory (Cambridge
University Press, New Delhi, 2010).
\end{thebibliography}
\end{document}